\documentclass[12pt]{article}
\usepackage{graphics}

\makeatletter

\providecommand{\LyX}{L\kern-.1667em\lower.25em\hbox{Y}\kern-.125emX\@}

\newcommand{\lyxaddress}[1]{
\par {\raggedright #1
\vspace{1.4em}
\noindent\par}
}

\newcommand\lesssim{\mathrel{\rlap{\lower4pt\hbox{\hskip1pt$\sim$}}
\raise1pt\hbox{$<$}}}
\newcommand\gtrsim{\mathrel{\rlap{\lower4pt\hbox{\hskip1pt$\sim$}}
\raise1pt\hbox{$>$}}}

\makeatother
\begin{document}

\title{Null Energy Condition Violations in Eternal Inflation%
\thanks{Talk given at COSMO-2001, Rovaniemi, Finland, August 29---September
4, 2001. At that time, the author's affiliation was with the Department
of Physics, Case Western Reserve University, 10900 Euclid Ave., Cleveland
OH 44106. Based on work in preparation \cite{GTW01}.
}}

\author{Serge Winitzki}

\maketitle

\lyxaddress{\centering Department of Physics and Astronomy, Tufts
University, Medford, MA 02155}

\begin{abstract}
The usual scenario of ``eternal inflation'' involves an approximately
de Sitter spacetime undergoing upward fluctuations of the local expansion
rate $H$. This spacetime requires frequent violations of the
Null Energy Condition (NEC). We investigate the fluctuations of the
energy-momentum tensor of the scalar field in de Sitter space as a
possible source of such violations. We find that fluctuations of the
energy-momentum tensor smeared in space and time are well-defined
and may provide the NEC violations. Our results for slow-roll inflation
are consistent with the standard calculations of inflationary density
fluctuations. In the diffusive regime where quantum fluctuations dominate
the slow-roll evolution, the magnitude of smeared energy-momentum
tensor fluctuations is large enough to create frequent NEC violations.
\end{abstract}
\begin{figure}
{\centering \resizebox*{2.4in}{!}{\includegraphics{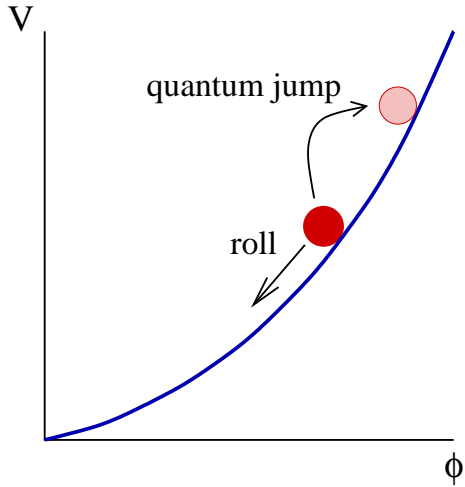}} \par}

\caption{\label{eternalidea} A scalar field rolls down the potential slowly
and causes inflation. The eternal inflation argument is that occasional
quantum jumps in localized regions can raise the scalar field to a
higher value of the energy density, leading to regions of faster inflation.
}
\end{figure}

The usual picture of eternal inflation \cite{earlywork} involves
a locally FRW spacetime which is well described by a de Sitter-like
metric with a slowly varying expansion rate $H$,
\begin{equation}
ds^{2}=dt^{2}-\left[ a\left( x,t\right) \right] ^{2}d\vec{x}^{2},\: a\left(
x,t\right) \approx e^{\int H\left( x,t\right) dt}.
\end{equation}
In a universe dominated by the vacuum energy $V\left( \phi \right) $
of the scalar field $\phi $ (the inflaton), the equations of
motion in the slow roll approximation are
\begin{equation}
\left( \frac{\dot{a}}{a}\right) ^{2}\equiv \left[ H\left( \phi \right)
\right] ^{2}=\frac{8\pi }{3}V\left( \phi \right) ,
\end{equation}
\begin{equation}
\label{eq:phi}
\dot{\phi }=-\frac{1}{4\pi }\frac{dH\left( \phi \right) }{d\phi }
\end{equation}
($G=c=1$ in our units). According to the standard picture, quantum
fluctuations of the inflaton on super-horizon scales $L\gtrsim H^{-1}$
behave classically and lead to a random variation of the field $\phi \left(
x,t\right) $ and of the expansion rate $H\left( x,t\right) $ on these
scales $L$. This modifies Eq.~(\ref{eq:phi}) so that, in addition
to the deterministic change, the field randomly jumps by $\delta \phi \sim
H/\left( 2\pi \right) $ in horizon-size regions during one Hubble time
$H^{-1}$ (see Fig.~\ref{eternalidea}). Occasionally, the field will move
upward on the potential, delaying the end of inflation. These random
``upward jumps'' create regions with a larger expansion rate.

For most inflaton potentials, there generically exists a regime where
the ``jumps'' significantly affect or even dominate the scalar
field evolution. The latter happens for values of $\phi $ such
that
\begin{equation}
\label{eq:fd}
\left| \dot{\phi }\right| \ll \frac{\left( H\left( \phi \right) \right)
^{2}}{2\pi }.
\end{equation}
If inflation starts with these values of $\phi $, the universe
will contain a statistical distribution of inflating regions, some
of which will thermalize; in that case inflation becomes ``eternal''
because at any (arbitrarily late) time there will be some regions
which have not yet reached thermalization. In this picture of eternal
inflation, upward jumps of the effective expansion rate $H$ in
horizon-size regions are assumed to be caused by backreaction of
fluctuations of the scalar field $\phi $ on the metric.

It has been known that eternal inflation requires violations of the
null energy condition \cite{BorVil}. The Einstein equation
\begin{equation}
\dot{H}=-4\pi G\left( \rho +p\right) +\frac{k}{a^{2}}
\end{equation}
shows that if $H$ is to increase and the curvature is zero or
negative ($k\leq 0$), the term $\rho +p$ must be negative.
For the standard energy-momentum tensor $T_{\mu \nu }$ of a homogeneous
fluid, this translates into the null energy condition (NEC),
\begin{equation}
\label{eq:tnn0}
T_{\mu \nu }N^{\mu }N^{\nu }<0
\end{equation}
for some null vector $N^{\mu }$. In other words, the NEC must
be violated in any patch of spacetime of size $L\gtrsim H^{-1}$
that undergoes an upward jump in its local expansion rate $H$.

The issue we are concerned with is the source of these NEC violations
during eternal inflation. Energy density of a classical scalar field
cannot violate the NEC, since

\begin{equation}
T_{\mu \nu }N^{\mu }N^{\nu }=\left( N^{\mu }\partial _{\mu }\phi \right)
^{2}\geq 0.
\end{equation}
A quantum field $\hat{\phi }$ may violate the NEC in some quantum
states; however, the expectation value of the (renormalized)
energy-momentum tensor $\hat{T}^{ren}_{\mu \nu }$ of a scalar field in a de
Sitter-invariant state does not cause violations of the NEC. In such a
state
\begin{equation}
\left\langle \hat{T}_{\mu \nu }^{ren}\right\rangle \propto g_{\mu \nu }
\end{equation}
because of de Sitter invariance, and therefore the NEC operator has
vanishing expectation value,
\begin{equation}
\left\langle \hat{O}^{ren}\right\rangle \equiv \left\langle \hat{T}_{\mu
\nu }^{ren}N^{\mu }N^{\nu }\right\rangle =0.
\end{equation}

Even if the expectation value of $\hat{T}_{\mu \nu }$ does not
violate the NEC, the energy-momentum tensor has fluctuations around
the mean which may be significant. In the picture of semiclassical
gravity, the spacetime metric is completely determined by the expectation
value of the energy-momentum tensor and is insensitive to its fluctuations.
We would like to include the effect of energy-momentum fluctuations
on the metric and for that we would need to go beyond semiclassical
gravity. In the present work we restrict ourselves to an evaluation
of $\langle \hat{O}^{2}\rangle $ and do not consider the effect
of the fluctuations on the metric, which is a separate and more complicated
issue.

A direct evaluation of the fluctuations of $\hat{T}_{\mu \nu }$
is problematic since the two-point function of the energy-momentum
tensor $\left\langle \hat{T}_{\mu \nu }\left( x\right) \hat{T}_{\rho \sigma
}\left( x\right) \right\rangle $ diverges at coincident points $x$. Even if
this divergence were somehow to be renormalized away, any remaining finite
piece will have to be of the form
\begin{equation}
\propto g_{\mu \nu }g_{\rho \sigma }+(\textrm{permutations})
\end{equation}
and, after a contraction with $N^{\mu }N^{\nu }N^{\rho }N^{\sigma }$,
will not contribute to the NEC.

From the formal point of view, the renormalized energy-momentum tensor
$\hat{T}_{\mu \nu }^{ren}$ and the NEC $\hat{O}^{ren}$ are
not operators but operator-valued distributions that become well-defined
operators after averaging with a window function, for instance
\begin{equation}
\label{eq:Tave0}
\hat{O}^{ren}_{W}\equiv \int d^{4}x\sqrt{-g}\, W\left( \frac{x}{L}\right)
\hat{O}^{ren}.
\end{equation}
Here the window function profile $W\left( x\right) $ is chosen
so that it falls off rapidly and provides a smearing on scale $L$.
Averaging over time as well as over space is necessary to avoid divergences
in $\langle (\hat{O}_{W}^{ren})^{2}\rangle $. (Details of the
calculations will be given in \cite{GTW01}). We shall adopt the viewpoint
that the quantum energy-momentum tensor smeared in space and time
on scales $L\gtrsim H^{-1}$ behaves quasi-classically and may
provide NEC violations in certain horizon-size regions. This is similar
to the assumption made in usual calculations of inflationary perturbations
where a coarse-graining on super-horizon scales is performed.

We take the quantum state of the field $\hat{\phi }$ during slow-roll
inflation to be a superposition
\begin{equation}
\hat{\phi }\left( x,t\right) =\phi _{0}\left( t\right) +\delta \hat{\phi
}\left( x,t\right) ,
\end{equation}
where $\phi _{0}\left( t\right) $ is the classical slow-roll
trajectory (represented by a suitable coherent state) and $\delta \hat{\phi
}$ is a quantum scalar field in the usual Bunch-Davies vacuum state of
the approximately de Sitter spacetime. We have computed the expectation
value $\left\langle \hat{O}\right\rangle $ and the dispersion
$\left\langle \hat{O}^{2}\right\rangle $ of the operator given
by Eq.~(\ref{eq:Tave0}) in this quantum state. Although the expectation
value $\left\langle \hat{O}\right\rangle >0$ in this state, the
NEC may be frequently violated if $\left\langle \hat{O}^{2}\right\rangle
\gg \left\langle \hat{O}\right\rangle ^{2}$.

The calculations were made for a massless field $\delta \hat{\phi }$
in de Sitter space, with a constant null vector field
\begin{equation}
N^{\mu }\propto \left( H\eta \right) ^{2}\left[ 1,{\mathbf n}\right] ,\;
{\mathbf n}=const,
\end{equation}
and an arbitrary normalized window profile $W\left( x\right) $
that decays quickly (it is sufficient to assume exponential decay)
for $\left| x\right| \gtrsim 1$. The smearing scales of interest
are $L=(\varepsilon H)^{-1}$ for both space and time (with $\varepsilon
\lesssim 1$). Under these assumptions we have obtained the ratio of
dispersion to mean,
\begin{equation}
\label{eq:ans}
\frac{\left\langle \hat{O}^{2}\right\rangle }{\left\langle
\hat{O}\right\rangle ^{2}}=\left( \frac{H^{2}}{2\pi \dot{\phi }_{0}}\right)
^{2}\max \left( c_{1}\varepsilon ^{2},c_{2}\varepsilon ^{4}\right) +\left(
\frac{H^{2}}{2\pi \dot{\phi }_{0}}\right) ^{4}c_{3}\varepsilon ^{8}
\end{equation}
where $c_{1}$, $c_{2}$, $c_{3} \sim O\left( 1\right) $
are window-dependent constants. (This result is otherwise insensitive
to the choice of the window profile.) The first term on the RHS of
Eq.~(\ref{eq:ans}) corresponds to first-order terms in the perturbative
expansion in $\delta \phi $, while the second term comes from
second-order expansion terms.

Within our assumptions, it follows from Eq.~(\ref{eq:ans}) that
fluctuations of the energy-momentum tensor lead to frequent NEC violations
when
\begin{equation}
\left| \dot{\phi }_{0}\right| \ll \frac{H^{2}}{2\pi }.
\end{equation}
Note that this is the same condition as Eq.~(\ref{eq:fd}) for fluctuation
domination. If this condition holds, the second term on the RHS of
Eq.~(\ref{eq:ans}) is of the same order or greater than the first
term; in other words, perturbation theory is inadequate in the
fluctuation-dominated regime. In the opposite regime, perturbation theory
is valid, the NEC holds and we obtain an agreement with standard
calculations of inflationary perturbations.

We have shown that there exists a range of scalar field values $\phi $
where fluctuations of the smeared energy-momentum tensor on super-horizon
scales are significant and suggest frequent NEC violations. This range
of $\phi $ coincides with the regime where the quantum ``jumps''
dominate the evolution of $\phi $, which is required for eternal
inflation. In the traditional description of inflation, the quantum
field $\phi $ smeared on super-horizon scales is treated as a
classical field. If super-horizon scale smearing is an accurate
phenomenological description of the quantum-to-classical transition of the
vacuum fluctuations in de Sitter space, these fluctuations would indeed
seem to provide the necessary NEC violations.

\begin{figure}
{\centering \resizebox*{2.4in}{!}{\includegraphics{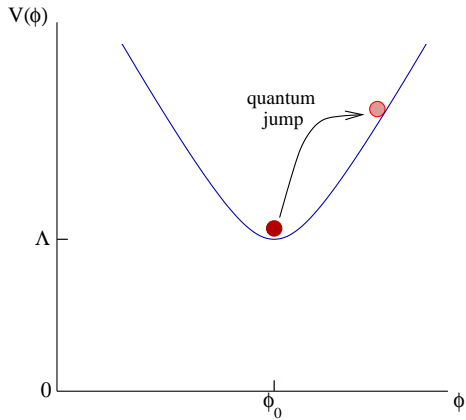}} \par}

\caption{\label{massfig} A classical scalar field with ground state at the
bottom of the potential with non-zero energy density $ \Lambda $
causing de Sitter expansion. Smeared quantum fluctuations of the
energy-momentum tensor may violate the NEC. In the heuristic picture, these
violations would correspond to localized quantum jumps of the scalar field
to higher values of the potential, causing faster inflation in those
regions and an instability of de Sitter expansion towards eternal
inflation.}
\end{figure}

Following the same approach, the present analysis suggests a possibility
of spontaneous eternal inflation in de Sitter spacetime. Suppose we
had a massive scalar field with the potential shown in Fig.~\ref{massfig}.
The classical field is located at the bottom of the potential and
gives rise to a background inflation. The quantum state of the fluctuations
of $\phi $ is assumed to be the Bunch-Davies vacuum. In this
state there will also be fluctuations of the energy density and, since
the expectation value $\left\langle \hat{O}\right\rangle =0$
while $\left\langle \hat{O}^{2}\right\rangle >0$, violations
of the NEC will happen. So the standard picture of eternal inflation
picture would suggest that de Sitter space is destabilized by the
presence of a scalar field in the Bunch-Davies vacuum.

One may also apply similar considerations to Minkowski spacetime and
conclude, by setting $\Lambda =0$ in Fig.~\ref{massfig}, that
Minkowski spacetime is unstable to eternal inflation. This conclusion
would be contrary to the usual assumption that flat spacetime is stable.
However, there is a crucial difference between Minkowski spacetime
and a de Sitter spacetime, namely that the ``horizon scale'' $H^{-1}$
in Minkowski spacetime is infinite. Therefore for Minkowski spacetime
to be unstable to eternal inflation, the NEC-violating fluctuation
would have to survive for an infinitely long time. The probability
for this to happen vanishes and hence Minkowski spacetime is stable
towards eternal inflation. However, in FRW spacetimes with a finite
horizon size we might still find an instability towards eternal inflation.
To resolve this question, a more precise picture of the backreaction
is needed.

An unsatisfactory feature of our argument is that it only concerns
itself with the spacetime before and after the fluctuation. The evolution
of the spacetime during the fluctuation itself is not considered.
Our attempt to go beyond the semiclassical analysis by calculating
fluctuations of the energy-momentum tensor was not entirely successful
since we do not yet have a scheme for calculating the backreaction
of NEC violating fluctuations on the metric. This issue requires further
study.

\end{document}